\documentclass[lettersize,journal]{IEEEtran}

\usepackage{tikz}
\usepackage{textcomp}
\usepackage{hyperref}
\usepackage{lipsum}

\usepackage{amsmath,amsfonts,amsthm,amssymb}
\usepackage{algorithmic}
\usepackage{algorithm}
\usepackage{url}
\usepackage{graphicx}
\usepackage{cite,xcolor}
\usepackage{bm}

\newtheorem{thm}{Theorem}[section]
\newtheorem{lem}[thm]{Lemma}

\theoremstyle{definition}
\newtheorem{defn}{Definition}[section]

\def\T{\mathsf{T}}    
\def\bxi{{\bm{\xi}}}
\def\x{{\bf x}}
\def\z{{\bf z}}
\def\u{{\bf u}}
\def\y{{\bf y}}
\def\f{{\bf f}}
\def\g{{\bf g}}
\def\I{{\bf I}}
\def\Lb{{\bf L}}
\def\A{{\bf A}}
\def\B{{\bf B}}
\def\C{{\bf C}}
\def\Q{{\bf Q}}
\def\K{{\bf K}}
\def\Pb{{\bf P}}
\def\R{{\bf R}}
\def\P{{\bf P}}
\def\0{{\bf 0}}
\def\H{{\bf H}}
\def\Yc{{\cal Y}}
\def\Rn{{\mathbb{R}}}
\def\L{{\cal L}}    
\def\Ls{{\mathfrak{L}}}    
\def\bXi{{\bm{\Xi}}}
\def\bR{{\bf R}}

\newcommand\copyrighttext{%
  \large\sf\textcopyright~2023 IEEE. Personal use of this material is permitted.
  Permission from IEEE must be obtained for all other uses, in any current or future
  media, including reprinting/republishing this material for advertising or promotional
  purposes, creating new collective works, for resale or redistribution to servers or
  lists, or reuse of any copyrighted component of this work in other works. 
  doi: \href{https://doi.org/10.1109/TIV.2023.3272660}{10.1109/TIV.2023.3272660}}
\newcommand\copyrightnotice{%
\begin{tikzpicture}[remember picture,overlay]
\node[anchor=south,yshift=100pt] at (current page.south) {\fbox{\parbox{\dimexpr\textwidth-\fboxsep-\fboxrule\relax}{\copyrighttext}}};
\end{tikzpicture}%
}

\begin{document}
\onecolumn

\copyrightnotice

\vfil

{\Large \noindent\sf K. Ampountolas, ``The Unscented Kalman Filter for Nonlinear Parameter Identification of Adaptive Cruise Control Systems," in IEEE Transactions on Intelligent Vehicles, doi: \href{https://doi.org/10.1109/TIV.2023.3272660}{10.1109/TIV.2023.3272660}.}

\bigskip \bigskip \bigskip \bigskip \bigskip 

{\Large \noindent\sf The material cannot be used for any other purpose without further permission of the publisher and is for private use only.}
\bigskip

\bigskip \bigskip \bigskip 
{\Large \noindent \sf
There may be differences between this version and the published version. You are advised to consult the publisher’s version if you wish to cite from it.}

\twocolumn
\clearpage

\title{The Unscented Kalman Filter for\\ Nonlinear Parameter Identification of\\ Adaptive Cruise Control Systems}

\author{Konstantinos Ampountolas,~\IEEEmembership{Member IEEE}
\thanks{This work was supported by the Center of Research, Innovation \& Excellence  (CRIE), University of Thessaly, under the project ``campaigningACC".}
\thanks{The author is with the Automatic Control \& Autonomous Systems Laboratory, Department of Mechanical Engineering, University of Thessaly, 38334 Volos, Greece (e-mail: k.ampountolas@uth.gr).}
}

\markboth{IEEE Transactions on Intelligent Vehicles}%
{Ampountolas: The Unscented Kalman Filter for Nonlinear Parameter Identification of Commercial ACC Systems}

\IEEEpubid{0000--0000/00\$00.00~\copyright~2023 IEEE}

\maketitle

\begin{abstract}
This paper develops and investigates a dual unscented Kalman filter (DUKF) for the   joint nonlinear state and parameter identification of commercial adaptive cruise control (ACC) systems. Although the core functionality of stock ACC systems, including their proprietary control logic and parameters, is not publicly available, this work considers a car-following scenario with a human-driven vehicle (leader) and an ACC engaged ego vehicle (follower) that employs a constant time-headway policy (CTHP). The objective of the DUKF is to determine the CTHP parameters of the ACC by using real-time observations of space-gap and relative velocity from the vehicle's onboard  sensors. Real-time parameter identification of stock ACC systems is essential for assessing their string stability,  large-scale deployment on motorways, and impact on traffic flow and throughput. In this regard, $\Ls_2$ and $\Ls_\infty$ string stability conditions are considered. The observability rank condition for nonlinear systems is adopted to evaluate the ability of the proposed estimation scheme to estimate stock ACC system parameters using empirical data. The proposed filter is evaluated using empirical data collected from the onboard sensors of two 2019 SUV vehicles, namely Hyundai Nexo and SsangYong Rexton,  equipped with stock ACC systems; and is compared with batch and recursive least-squares optimization. The set of ACC model parameters obtained from the proposed filter revealed that the commercially implemented ACC system of the considered vehicle (Hyundai Nexo) is neither $\Ls_2$ nor $\Ls_\infty$ string stable. 
\end{abstract}


\section{Introduction}

\IEEEPARstart{A}{daptive cruise control} (ACC) systems, which belong to Level-1 driving automation of the Society of Automotive Engineers (SAE)  \cite{SAE:2021}, are already available as optional or standard equipment in commercially available vehicles. In car-following or platooning scenarios with a human-driven vehicle (leader) and a number of ACC-equipped ego vehicles (followers), the ACC system controls the longitudinal motion of the equipped vehicles by observing the velocity and distance from the leader to track a user-defined time headway or reference velocity. To achieve this goal, the ACC system adjusts the ego vehicle’s velocity by accelerating or decelerating it. 

Automation of the longitudinal movement of vehicles in platoons unveils two fundamental aspects of the ACC system: (a) the {\it spacing policy} (or controller), which specifies the user-defined desired inter-vehicular distance (time gap or space gap); and (b) the {\it string stability} of the platoon in the presence of disturbancies \cite{Peppard:1974}.   Several different spacing policies have been proposed for ACC systems, of which the {\it constant spacing policy} \cite{Swaroop:spacing}, the {\it constant time headway policy} \cite{Ioannou:1993}, and the {\it variable time headway policy} \cite{Kanellakopoulos:VTH} are the most remarkable. A comparison can be found in \cite{doi:10.1080/00423119408969077}.

String stability of interconnected systems (e.g., car-following or platooning systems; spring-mass systems, irrigation systems) has been a central topic of research in the control community for decades \cite{Peppard:1974,Swaroop:1996,Eyre:1998,doi:10.1076/vesd.32.4.313.2083,983389,Ploeg:2014,Besselink}. String stability in car-following formation characterizes the upstream amplification of random perturbations through the platoon of vehicles. Recently, the assessment of commercially implemented ACC systems in car-following scenarios using empirical observation has been shown to be {\it string unstable} in the $\Ls_2$ or $\Ls_\infty$ sense\cite{MILANES2014,Knoop,doi:10.1177/0361198120911047,MAKRIDIS2021103047,Gunter:2020,Gunter:2021,Wang:2021}.

\IEEEpubidadjcol 

Real-time parameter identification of commercially implemented ACC systems using empirical observations is essential to assess the string stability of ACC systems; their deployment at scale on motorways; and their impact to traffic flow and throughput. However, the parameter identification of nonlinear systems, such as the stock ACC system of automated vehicles, using empirical observations is challenging since the underlying observability problem is non-convex and it might be ill-conditioned under equilibrium driving conditions (i.e., where acceleration and space-gap reduce to zero),  as shown in \cite{Wang:2021}. In the latter case, the ACC system parameters cannot be uniquely identified, given input and output observations from the platoon.

The present work develops and investigates a dual unscented Kalman filter (DUKF) \cite{Julier:AC_2000,Wan:ch7,Julier_UKF} for the nonlinear joint state and parameter identification of commercially implemented ACC systems that employ a constant time-headway policy (CTHP), unlike previous works where batch optimization, recursive least-squares, and particle filtering are used \cite{Punzo:Calib,Kesting:Calib,Wang:2021}.  The parameter identification problem of automated vehicles can be also tackled using surrogate models (see e.g.,  the Gaussian process-based model proposed in \cite{GaussianParameter:2022}) to approximate the longitudinal movement of automated vehicles in platoons and their (unknown) stock ACC system or other advanced driver assistance system (ADAS). Surrogate models can be trained to learn the personalized driving behavior of drivers using off-line or real-time data, and thus to design ADAS suitable to driver's preferences. In this regard, the  proposed DUKF aims at determining the CTHP parameters of commercial ACC systems (or other ADAS), given real-time observations of space headway and relative velocity from on-board vehicle sensors.

Observability describes the possibility of inferring the system state by observing its inputs and outputs.  To assess the ability to identify the ACC system parameters via the proposed DUKF or other nonlinear filtering approaches, the present paper employs an analytic algebraic condition, the so-called {\it observability rank condition} (ORC), for the determination of the observability of nonlinear systems \cite{Hermann:1977,NIJMEIJER:1982}. This is in contrast to the observability rank criterion of linear systems, which has been used in other works \cite{Wang:2021}.

The unscented Kalman filter (UKF), which can deliver better results as compared to the extended Kalman filter (EKF) \cite{Jazwinski} or other filters employing linearization \cite{Sarkka:2013}, is based on the {\it unscented transform} (UT).  The UT {\it deterministically} chooses a number of {\it sigma points} that estimate the mean and covariance of the probability distribution of the physical system state. These sigma points are then plugged into the nonlinear operator of the measurements model, and the mean and covariance of the output are estimated from them. 

Although the UT resembles Monte Carlo estimation algorithms (e.g., particle filtering), the approaches are different. Since the {\it sigma points are selected deterministically} in the UT. The UKF is not based on Taylor series-based local approximations (e.g., linear or quadratic) at a single point such as the EKF, but uses further points in approximating the nonlinearity of the state and measurement model \cite{Julier_UKF,Sarkka:2013}.  However, the UKF requires slightly more computational operations than the EKF, while it requires less computational effort than particle filters. Concluding, the UKF is suitable for nonlinear systems and thus favourable for the real-time system identification of the ACC model parameters.

\emph{Statement of Contributions:} The present work: 
(a) It develops a {\it dual unscented Kalman filter} for the nonlinear state and parameter identification of commercial  ACC systems  that employ a {\it constant time-headway policy}. (b) It presents the {\it observability rank condition} for the determination of the observability of nonlinear systems, unlike previous works that employed the ORC for linear systems. This condition provides  insights on the ability to estimate stock ACC model parameters using empirical data though nonlinear filtering approaches. (c) It demonstrates that the set of ACC model parameters obtained from the proposed estimation scheme using empirical data reveal that the ACC system of a stock 2019 SUV is neither $\Ls_2$ nor $\Ls_\infty$ string stable.

\emph{Organization:} Section \ref{sec:ACC} reviews the constant time-headway policy for ACC systems and its stability conditions, and presents the nonlinear parameter identification problem of  ACC systems.  Section \ref{sec:NonLinORC} presents and applies the observability rank condition for the determination of the observability of the ACC system.  Section \ref{sec:UKF} presents the proposed dual unscented Kalman filter for the nonlinear state and parameter identification of stock ACC systems. Section \ref{sec:results} demonstrates the efficacy of the DUKF (and its comparison to least-squares optimization) using empirical data collected from a car-following scenario involving two 2019 model year SUV vehicles equipped with stock ACC systems. Finally, Section \ref{sec:conclusions} provides research directions for future work.

{\it Notation}: The fields of real and complex numbers are denoted by $\Rn$ and $\mathbb{C}$, respectively. The imaginary unit is denoted by $j$, where $j := \sqrt{-1}$. The space of Lebesgue measurable functions $f: \Rn_+ \to \Rn$ such that $t \to |f (t)|^{\mathsf{p}}$ is integrable over $\Rn$ is denoted by $\Ls_{\mathsf{p}}$, here $\mathsf{p} =   2, \infty$ is used to discuss string stability. For $\mathsf{p} = \infty$ no integration is used, and instead, the norm on $\Ls_\infty$ is given by the essential supremum. Given a transfer function $H(j\omega)$, $\omega\in\Rn$, of a single-input single-output (SISO) system, the ${\cal H}_\infty$ norm of the system is defined as $\| H \|_{{\cal H}_\infty} = \sup_{\omega\in\Rn} |H(j\omega)|$.

Given a scalar field $h(\bxi)$, with $\bxi \in \Rn^n$, $dh$ is its differential. Given a vector field $\f \in \Rn^n$, $\L_\f$ denotes the {\it Lie derivative} along $\f$. The Lie derivative along $\f$ of a given scalar field $h$ is $\L_{\f} h = \partial h / \partial \bxi \cdot \f$; moreover it holds $\L_\f dh = d\L_\f h$. The $\kappa$-th Lie derivative  along $\f$ of a given scalar field $h$ is $\L_{\f}^{(\kappa)}h = \underbrace{\L_{\f}\L_\f \cdots \L_\f}_{\kappa \text{ times}} h$  \cite{Isidori:1995}.

\section{Design of ACC Systems}\label{sec:ACC}

\subsection{ACC with Constant Time-Headway Policy}

Consider the  {\it constant time-headway policy} (CTHP) for a car-following scenario with a human-driven vehicle (leader) and an ACC engaged ego vehicle (follower):
\begin{align}\label{eq:p}
	\dot{p}(t) &= \Delta v(t), \\ \label{eq:CTHP}
	\dot{v}(t) &= \alpha \left[p(t) - \tau v(t) \right] + \beta \Delta v(t) + d(t),
\end{align}
where $p(t) \triangleq p_l(t) - p_f(t) - L$ is the space gap between the two vehicles, with $p_l$ [m] the position of the leader, $p_f$ [m] the position of the follower and $L$ [m] the length of leading vehicle; $v(t)$ [m/s] and $\Delta v(t) = u(t) - v(t)$ [m/s] are the velocity of the ACC ego vehicle and the velocity difference between the leading vehicle and ACC vehicle, respectively. The term $\delta \triangleq \tau v(t)$ [m] characterizes the user-defined space gap, parameterized by the desired constant-time headway, $\tau$ [s], that the ACC aims to maintain.  The term $d$ captures the effect of external disturbances on acceleration, but they might result from modeling errors or parameter uncertainties. The two non-negative control gains $\alpha$ [1/s$^2$] and $\beta$ [1/s] control the constant-time headway and the relative velocity terms, respectively. The parameter $\tau$ [s] is the time-gap at equilibrium. The two control gains should be selected such that the eigenvalues of the associated closed-loop system have negative real parts. 

It is assumed that the lead vehicle velocity, $u(t)$, is available (measured using range sensors) in real-time (ACC scenario) or that the lead vehicle broadcasts its velocity to the ACC ego vehicle via vehicle-to-vehicle communication (cooperative ACC (CACC) scenario), thus the ACC ego vehicle can implement the controller \eqref{eq:CTHP} in real-time. In this implementation, the lead vehicle chooses its own input (e.g., acceleration/deceleration) without regard to the follower, and the ACC engaged ego vehicle applies the controller \eqref{eq:CTHP} to automatically follow the leader with desired constant-time headway $\tau$.


\subsection{$\Ls_2$ and $\Ls_\infty$ String Stability}\label{sec:stability}

This section presents the stability properties of the control policy \eqref{eq:CTHP} concerning its (unknown) design parameters $\alpha$, $\beta$ and $\tau$ (see \cite{Sheikholeslam:1993,Wilson_stability,L2Linf_stab} for details). Stability analysis of automated vehicles in car-following formation rely on notions of string stability. It characterizes the upstream amplification of random disturbances through the platoon of vehicles. $\Ls_2$ string stability refers to the energy or variance dissipation of the output signal. $\Ls_\infty$ string stability has a physical meaning as it concerns with the amplitude of the signal deviations, and thus it can be directly related to a qualification for collision avoidance, traffic safety, and traffic flow throughput.

{\it 1) $\Ls_2$ Strict String Stability:} Considering input-output stability, a sufficient condition for $\Ls_2$ {\it strict string stability} is as follows \cite{Sheikholeslam:1993}:
\begin{equation}\label{eq:L2}
	|H(j\omega)| = \sqrt{\frac{\alpha^2 +\beta^2 \omega^2}{(\alpha-\omega^2)^2+\omega^2(\alpha \tau +  \beta)^2}} \le 1, \; \forall \, \omega \ge 0,
\end{equation}
where $H(j\omega)$ is the speed-to-speed (or headway-to-headway) transfer function, 
\begin{equation}\label{eq:tf}
	H(s) = \frac{\beta s + \alpha}{s^2 + (\alpha \tau +  \beta)s + \alpha},
\end{equation}
evaluated at $s := j\omega \in \mathbb{C}$ and $\omega \ge 0$ is the frequency. The sufficient condition \eqref{eq:L2} leads to the following condition on the  ACC model parameters $\alpha$, $\beta$ and $\tau$: $$\alpha^2 \tau^2 + 2\alpha \beta \tau - 2\alpha \ge 0.$$ Note that as $\tau$ approaches $\infty$ the system is $\Ls_2$ strict string stable for all non-negative gains $\alpha$ and $\beta$ of the CTHP (actually, it becomes independent of $\beta$, and thus of the leader's velocity), while for small values of the time gap (i.e., as $\tau$ approaches zero) the system becomes unstable.

{\it 2) $\Ls_\infty$ Strict String Stability:} A sufficient condition for $\Ls_\infty$ strict string stability is to have the $\Ls_1$ norm of the impulse response less than 1. Note that ${\cal H}_\infty$  of a transfer function (peak value of $|H(j\omega)|$) is finite if and only if the transfer function is stable (otherwise, it is infinite). Moreover, the following holds: (a) the ${\cal H}_\infty$ norm is upper bounded by the $\Ls_\infty$-induced norm; (b) the two norms are identical for nonnegative impulse responses \cite{Boyd:1991}. A sufficient condition for obtaining a monotonic step response is nonimaginary poles and negative zeros in the transfer function \eqref{eq:tf}, which yields:
\begin{align*}
(\alpha \tau +  \beta)^2 - 4\alpha  \ge 0\,\, \text{ and }\,\, \alpha/\beta > 0.
\end{align*}
Obviously, the last condition is always satisfied due to the physical meaning of the nonnegative  gains $\alpha$ and $\beta$.

In summary, for the control policy \eqref{eq:CTHP} to be $\Ls_2$ and $\Ls_\infty$ {\it strict string stable}, the following conditions must be satisfied \cite{Wilson_stability,L2Linf_stab}:
\begin{equation}\label{eq:L2c}
\alpha^2 \tau^2 + 2\alpha \beta \tau - 2\alpha \ge 0,
\end{equation}
and
\begin{equation}\label{eq:Linfc}
(\alpha \tau +  \beta)^2 - 4\alpha \ge 0,
\end{equation}
respectively. Moreover, substracting \eqref{eq:L2c} from \eqref{eq:Linfc} yields \cite{L2Linf_stab}: 
\begin{equation}
\beta^2 \ge  2\alpha \;\; \Rightarrow\;\; \left(\text{$\Ls_\infty$ stability $\Leftrightarrow$ $\Ls_2$ stability}\right),
\end{equation}
suggesting that  $\Ls_2$ stability is stronger (i.e., more conservative) than the $\Ls_\infty$ stability. This is realistic since even if the $\Ls_2$ norm of a signal (energy dissipation) is small, it may occasionally contain large peaks, provided the peaks (i.e., the $\Ls_\infty$ norm) are not too frequent and do not contain too much energy. 

Section \ref{sec:results} shows that the set of ACC model parameters obtained from the proposed DUKF using empirical data reveals that the commercially implemented ACC system of a stock 2019 SUV is neither $\Ls_2$ nor $\Ls_\infty$ strict string stable, see also \cite{MILANES2014,Gunter:2020,Gunter:2021,Wang:2021} for more insights.

\subsection{Nonlinear Parameter Identification of ACC Systems}

The goal is to develop a nonlinear dual filtering approach that  simultaneously delivers one-step predictions for the states $p$ and $v$ and real-time estimates for the constant but unknown parameters $\alpha$, $\beta$ and $\tau$ that characterize the ACC system. 

Let the vector of the CTHP parameters be $\bm{\theta} = [ \alpha\,\, \beta\,\, \tau]^\T$. Then the (noise-free) continuous-time system \eqref{eq:p}--\eqref{eq:CTHP} is rewritten in discrete-time using Euler's forward discretization scheme with sampling time $T$ and introduces the additional state equation $\bm{\theta}_{k+1} = \bm{\theta}_k$ since the ACC model parameters are assumed to remain constant in time:
\begin{equation}\label{eq:discrete_matrix}
\begin{bmatrix}
		p_{k+1} \\
		v_{k+1} \\
		\alpha_{k+1} \\
		\beta_{k+1}\\
		\tau_{k+1}
	\end{bmatrix}
	= \begin{bmatrix}
		p_k + T (u_k - v_k) \\
		v_k + T \big[\alpha_k (p_k - \tau_k v_k ) + \beta_k ( u_k - v_k )\big]\\ 
		\alpha_k \\
		\beta_k \\
		\tau_k
	\end{bmatrix}.
\end{equation}
The augmented system state reads $\bxi_k= \left[ p_k\,\, v_k\,\, \alpha_k\,\, \beta_k\,\, \tau_k\right]^\T \in \Rn^5$. The model \eqref{eq:discrete_matrix} can then be re-written in compact vector form as $\bxi_{k+1} = \f(\bxi_k, u_k)$, where $\f \in\Rn^5$ is a nonlinear vector function reflecting the right-hand side of \eqref{eq:discrete_matrix}. The nonlinearity here appears due to the product of the physical states $p$ and $v$ with the CTHP parameters of the ACC system $\alpha$, $\beta$, and $\tau$. Finally, the measurement equation that reflects the physical system is given by:
\begin{equation}\label{eq:meas}
\y_k = {\bf C } \bxi_k = \begin{bmatrix}
1 & 0 & 0 & 0 & 0\\
0 & 1 & 0 & 0 & 0
\end{bmatrix} \bxi_k = \begin{bmatrix}
p_{k} \\
v_{k}
\end{bmatrix},
\end{equation}
where $\y_k\in \Rn^2$ is the measurement vector, and ${\bf C}$ is the corresponding state measurement matrix.

It should be noted that alternative discretization schemes may be employed in \eqref{eq:discrete_matrix} (Euler's backward method, Runge--Kutta method, Adams-Bashforth or Adams-Moulton methods, see e.g., \cite{BALDI201681}.) to provide a better approximation of the continuous-time dynamics in \eqref{eq:p}--\eqref{eq:CTHP}. This is important for any nonlinear estimation scheme since the  observability rank condition for nonlinear systems presented in Section \ref{sec:NonLinORC} is  susceptible to the type of the adopted discretization scheme.

\section{Nonlinear Observability Analysis}\label{sec:NonLinORC}

This section employs an analytic algebraic condition, the so-called {\it observability rank condition} (ORC), for the determination of the observability of the nonlinear system \eqref{eq:discrete_matrix}--\eqref{eq:meas}.  The ORC provides some insights on the ability to estimate the CTHP parameters of the ACC system, $\alpha$, $\beta$ and $\tau$, via the DUKF (see Section \ref{sec:UKF}) or other nonlinear filtering approaches.

Consider the following noise-free nonlinear state and measurement equations describing \eqref{eq:discrete_matrix}--\eqref{eq:meas}:
\begin{subequations}\label{eq:Nonlinear_Noise_free}
\begin{align}\label{eq:NL1}
	\bxi_{k+1} &= \f_k\left(\bxi_k, \u_k\right), \\ \label{eq:NL2}
	\y_k &= \g\left(\bxi_k,\u_k\right).
\end{align}
\end{subequations}
where $\bxi\in\Rn^n$, $\u\in\Rn^q$, and $\y\in\Rn^l$ are the state, control, and measurements vectors, respectively; and  $\f$, $\g$ are nonlinear vector functions of appropriate dimension. Moreover, consider the availability of the following trajectories:
\begin{enumerate}
	\item A state $\bxi^{(i)} \in \Rn^n$.
	\item An admissible control trajectory up to time $\kappa = k-1$: ${\cal U}_{\kappa}^{(i)} := \{\u_1^{(i)}, \u_2^{(i)}, \ldots, \u_\kappa^{(i)}\}$ (e.g., the measured using range sensors lead vehicle velocity, $u_\kappa$).
	\item An output measurements trajectory up to time $\kappa = k-1$: $\Yc_{\kappa}^{(i)} := \{\y_1^{(i)}, \y_2^{(i)}, \ldots, \y_\kappa^{(i)}\}$ that correspond to the state $\bxi^{(i)}$, for a given admissible control  ${\cal U}_{\kappa}^{(i)}$.
\end{enumerate}

The following definitions of observability will help us to define the observability of nonlinear systems \cite{Hermann:1977,NIJMEIJER:1982,LEE199149}.
\begin{defn}[Indistinguishable States]
Two states, $\bxi^1$ and $\bxi^2$ are {\it indistinguishable}, denoted as $\bxi^1 {\cal I} \bxi^2$, if they yield identical outputs for all admissible control inputs, that is, if $\y^{1}_k = \y^{2}_k$, for all $k$.
\end{defn}

\begin{defn}[Observability] 
The nonlinear system \eqref{eq:Nonlinear_Noise_free} is {\it observable} at $\bxi$, if the set of states that are indistinguishable from $\bxi$ includes only $\bxi$, that is, if ${\cal I}(\bxi) = \left\{ \bxi\right\}$. The system \eqref{eq:Nonlinear_Noise_free} is {\it observable} if all of its states are featured with this property. 
\end{defn}

\begin{defn}[Strong Observability] 
The nonlinear system \eqref{eq:Nonlinear_Noise_free}  is {\it strongly observable} at (or finite-time observable at) $\bxi^1$, if for all $\bxi^2$ and any admissible control input, $\y^1_k = \y^2_k$ implies that $\bxi^1 = \bxi^2$. The system  \eqref{eq:Nonlinear_Noise_free}  is {\it strongly observable} if all of its states are featured with this property.
\end{defn}

\begin{defn}[Strong Local Observability] 
The nonlinear system \eqref{eq:Nonlinear_Noise_free}  is {\it strongly locally observable} at $\bxi^1$, if there exists a neighborhood $\cal N$ of $\bxi^1$, such that for all $\bxi^2 \in \cal N$ and any admissible control input, $\y^1_k = \y^2_k$ implies that $\bxi^1 = \bxi^2$. The system \eqref{eq:Nonlinear_Noise_free}  is {\it strongly locally observable} if all of its states are featured with this property.
\end{defn}

The following lemma offers an algebraic observability condition for nonlinear systems \cite{Hermann:1977,LEE199149}.
\begin{lem}[Observability Rank Condition]\label{lem:ORC}
The nonlinear system \eqref{eq:Nonlinear_Noise_free} is {\it strongly locally observable} at state $\bm{\xi}$, if there exists a neighborhood of $\bm{\xi}$ and an $r$-tuple of integers $\kappa_1, \kappa_2, \ldots, \kappa_r$  with $\kappa_1 \ge \kappa_2 \ge \cdots \ge \kappa_r \ge 0$ and $\sum_{i=1}^r \kappa_i = n$, such that the following {\it observability matrix}, ${\cal O}_r$,  has rank $n$:
\begin{equation}\label{eq:ORC}
{\cal O}_r = \begin{bmatrix}
\tilde{{\cal O}}_1\\\tilde{{\cal O}}_2\\ \vdots\\ \tilde{{\cal O}}_r 
\end{bmatrix},\,
\tilde{{\cal O}}_i = \frac{\partial}{\partial \bm{\xi}}  
\begin{bmatrix}
d g_i(\bxi)\\
\L_\f^{(1)} g_i(\bxi)\\
\vdots\\
\L_\f^{(\kappa_i-1)} g_i(\bxi)\\
\end{bmatrix},\, i = 1, \ldots, r.
\end{equation}
If this condition holds, it is said that the nonlinear system fulfills the {\it observability rank condition} (ORC).
\end{lem}

Note that the only requirement on the individual $\kappa_i$ in Lemma \ref{lem:ORC} is that they sum to $n$. This implies that ${\cal O}_r$ is, in general, {\it not unique}. Moreover, for affine-input systems, {\it nonlinear observability is affected by the input}, though the control input (leader's velocity) is assumed to be known in our setting. This is in contrast to the observability rank criterion of linear systems (assume the standard $(\A, \B, \C)$ state-space form), where the observability matrix is only affected by the state and output matrices, respectively, $\A$ and $\C$, and not by the input matrix $\B$. Finally, applying \eqref{eq:ORC} for nonlinear systems on linear systems leads to the well-known linear observability rank criterion.

The ORC for the nonlinear system \eqref{eq:discrete_matrix}--\eqref{eq:meas} can  now be determined with $\bxi = \left[ p\,\,\, v\,\,\, \alpha\,\,\, \beta\,\,\, \tau\right]^\T$, $n = 5$, $q = 1$, $r = 2$, and $\kappa_1 = 3$, $\kappa_2 = 2$ (notice  $\kappa_1+\kappa_2=n$). The observability matrix can be then given by:
\begin{equation*}
{\cal O}_2 = \begin{bmatrix}
\tilde{{\cal O}}_1\\ \text{-- --}\\\tilde{{\cal O}}_2\end{bmatrix} =  \frac{\partial}{\partial \bm{\xi}}\begin{bmatrix}
dg_{1}(\bxi)\\ \L_\f^{(1)} g_1(\bxi)\\ \L_\f^{(2)} g_1(\bxi)\\  \text{-- -- -- --} \\dg_{2}(\bxi)\\ \L_\f^{(1)} g_2(\bxi)
\end{bmatrix} \in \Rn^{5 \times 5}.
\end{equation*}
The two elements of ${\cal O}_2$ can be calculated as follows:
\begin{itemize}
\item For determining $\tilde{{\cal O}}_1$ with $\kappa_1 = 3$:
\begin{align*}
&dg_{1}(\bxi) = \left[ 1\,\,\, 0\,\,\, 0\,\,\, 0\,\,\, 0\right]\begin{bmatrix}
p\\ v\\ \alpha\\ \beta\\ \tau\end{bmatrix}= p, \\ 
&\L_\f^{(1)} g_1(\bxi) = f_1(\bxi),\\ 
&\L_\f^{(2)} g_1(\bxi) = 
\left[ 1\,\,\, -T\,\,\, 0\,\,\, 0\,\,\, 0\right]
\begin{bmatrix}
f_1(\bxi)\\ f_2(\bxi)\\ f_3(\bxi)\\ f_4(\bxi)\\ f_5(\bxi)
\end{bmatrix}  = f_1(\bxi) - T f_2(\bxi).
\end{align*}
\item For  determining  $\tilde{{\cal O}}_2$ with $\kappa_2 = 2$:
\begin{align*}
&dg_{2}(\bxi) = \left[ 0\,\,\, 1\,\,\, 0\,\,\, 0\,\,\, 0\right]\begin{bmatrix}
p\\ v\\ \alpha\\ \beta\\ \tau\end{bmatrix} = v,\\  
&\L_\f^{(1)} g_2(\bxi) = f_2(\bxi).
\end{align*}
\end{itemize} 
From above, the observability matrix takes the final form:
\begin{equation*}
{\cal O}_2 = \begin{bmatrix} 1 & 0 & 0 & 0 &0\\
1 & -T & 0 & 0 & 0\\
\mathfrak{A} &  \mathfrak{B} & \mathfrak{C} & \mathfrak{D} & \mathfrak{E}\\
0 & 1 & 0 & 0 &0\\
\mathfrak{F} & \mathfrak{G} & \mathfrak{H} & \mathfrak{I} & \mathfrak{J}\\
\end{bmatrix},
\end{equation*}
with
\begin{align*}
&\mathfrak{A} = 1-\alpha T^2,
&\mathfrak{B} &= -2T+(\tau \alpha + \beta)T^2,\\
&\mathfrak{C} = -(p - \tau v )T^2, 
&\mathfrak{D} &=- ( u - v )T^2,\\
&\mathfrak{E} = \alpha vT^2,
&\mathfrak{F} &= \alpha T,\\
&\mathfrak{G}= -\alpha \tau T,
&\mathfrak{H}&= (p - \tau v )T,\\
&\mathfrak{I} = ( u - v )T,
&\mathfrak{J}&=-\alpha vT.
\end{align*}

The ORC is first investigated for equilibrium driving conditions where acceleration and space-gap reduces to zero, i.e.,
\begin{equation*}
u_k-v_k = 0\quad \text{and}\quad p_k - \tau_k v_k = 0.
\end{equation*}
Under this condition, the pairs $(\mathfrak{C}, \mathfrak{H})$ and $(\mathfrak{D}, \mathfrak{I})$ are zero, and the resulting ORC matrix has ${\rm rank}({\cal O}_2) = 3 \neq n = 5$, indicating a non-observable system. The Sylvester's law of nullity  (or rank-nullity theorem) suggests that ${\rm nullity}({\cal O}_2) = n - {\rm rank}({\cal O}_2) = 2$ and the corresponding null space (kernel) under equilibrium conditions is given by:
\begin{equation*}
{\rm null}({\cal O}_2) = \begin{bmatrix}
0 & 0 & 1 & 0 & 0\\
0 & 0 & 0 & 1 & 0
\end{bmatrix}^\T.
\end{equation*}

For non-equilibrium driving conditions, the pairs $(\mathfrak{C}, \mathfrak{H})$ or $(\mathfrak{D}, \mathfrak{I})$ are nonzero for any value of the involved quantities. Unfortunately, the resulting ORC matrix has ${\rm rank}({\cal O}_2) = 3 \neq n = 5$, indicating again a non-observable system with ${\rm nullity}({\cal O}_2) = n - {\rm rank}({\cal O}_2) = 2$. In this case, the corresponding null space is given by:
\begin{equation*}
{\rm null}({\cal O}_2) = \begin{bmatrix}
0 & 0 & -(u - v)/(p - \tau v) & 1 & 0\\
0 & 0 & (\alpha v)/(p - \tau v) & 0 & 1
\end{bmatrix}^\T.
\end{equation*}
Thus two parameters out of three of the ACC system cannot be identified at both equilibrium and non-equilibrium driving conditions by conventional filtering techniques which are based on linearization (e.g., EKF) or other nonlinear approaches.

Concluding, the application of the algebraic ORC of Lemma \ref{lem:ORC} suggests that the nonlinear system \eqref{eq:discrete_matrix}--\eqref{eq:meas} is non-observable for both equilibrium and non-equilibrium driving conditions. However, since {\it nonlinear observability is affected by the input} (i.e., the lead vehicle velocity) the ORC can be fulfilled (guarantee exponential convergence of the parameter error vector to zero) if the reference signal is {\it rich enough} (i.e., contain a certain or sufficient  number of frequencies) and satisfies an appropriate persistent excitation (PE) condition as in the model reference adaptive control (MRAC) \cite{BOYD1986_PE}. Notice also that the ORC is prone to the type of the employed discretization scheme.

\section{The Unscented Kalman Filter}\label{sec:UKF}

Consider the following nonlinear state and measurement models corrupted by noise:
\begin{subequations}\label{eq:Nonlinear_Noise}
\begin{align}\label{eq:Nonlinear_Noise_f}
	\bxi_{k} &= \f\left(\bxi_{k-1}, \u_{k-1}\right) + \bm{\gamma}_{k-1},\\ \label{eq:Nonlinear_Noise_g}
	\y_k &= \g_k(\bxi_k) + \bm{\zeta}_k,
\end{align}
\end{subequations}
where $\bxi \in \Rn^n$ is the system state, $\u \in \Rn^q$ is the control, and $\y \in \Rn^l$ is 
the output. The process noise $\bm{\gamma}$ and measurement noise $\bm{\zeta}$ are mutually independent (though they might be colored), zero-mean white Gaussian processes with covariances $\Q_{k-1}$ and $\R_k$, respectively. The nonlinear vector functions $\f: \Rn^n\times \Rn^q \to \Rn^n$ and $\g: \Rn^n \to \Rn^l$ represent the physical model dynamics and the observation model, respectively. 

The main goal is to estimate $\bxi_k$ from measurements of the output $\y_k$. Denote the output measurements up to time $\kappa$ as: $\Yc_\kappa := \{\y_1, \y_2, \ldots, \y_\kappa\}$. The state estimation problem then is to build an estimate of $\bxi_k$ using $\Yc_\kappa$ at each  $k>\kappa$.

The UKF estimation scheme is expressed at each measurement step $k=1, 2, 3, \ldots$ as  \emph{state prediction} and \emph{state update}.\\
{\it 1) Prediction at $k-1$ for $\bxi_{k-1}$}. Given the state vector $\bxi_{k-1}$ at step $k-1$ with a mean value $\hat{\bxi}_{k-1}$ and covariance  ${\bf P}_{k-1}^{\bxi}$,  the statistics of $\bxi$ are calculated by using the unscented transform (the sigma points and corresponding weights). Let $S_p = 2n + 1$ be the number of sigma points stacked in the vector $\bXi^{(i)} \in \Rn^n$, $i = 0, 1, \ldots, 2n$, where $n$ is the dimension of the system state $\bxi$, with corresponding weights $w^{(i)} \in (0,1)$, $i = 0, 1, \ldots, 2n$. The sigma points are then determined as follows:
\begin{subequations}\label{eq:Sigma_points}
\begin{align} 
\bXi^{(0)}_{k-1} &= \hat{\bxi}_{k-1} \in \Rn^n, \\ \label{eq:Sigma_pointsx1}
\bXi^{(i)}_{k-1} &= \hat{\bxi}_{k-1} + \delta \Big[\sqrt{{\bf P}_{k-1}} \Big]_i, \quad i = 1, \ldots, n,\\ \label{eq:Sigma_pointsx2}
\bXi^{(i+n)}_{k-1} &= \hat{\bxi}_{k-1} - \delta\Big[\sqrt{{\bf P}_{k-1}} \Big]_i, \quad i = 1, \ldots, n,
\end{align}
\end{subequations}
where $[\cdot]_i$ denotes the $i$-th column of the matrix square root of ${\bf P}$, and $\delta = \sqrt{n+\lambda}$. The scaling parameter, $\lambda = a^2(n+b)-n$, where the parameters $10^{-4}\le a \le 1$ (a small positive value) and $b$ (usually set to $0$ or $3-n$) controls the spread of the sigma points around the mean \cite{Julier:AC_2000,Wan:ch7}. For  any symmetric prior distribution with kurtosis $\mathsf{k}$ the selection of $b=\mathsf{k}-n$ allows for more accurate predictions of mean and covariance than those made by the EKF (which is based on linearization), while for  $\mathsf{k} =3$ the error in the kurtosis is minimized \cite{Julier:AC_2000}. Since the matrix $\Pb$ is positive definite it can be decomposed into $\Lb\Lb^\T$ via lower-triangular Cholesky factorization (note that $\Pb$ and $\Lb$ have the same eigenvectors). As can be seen in \eqref{eq:Sigma_pointsx1}--\eqref{eq:Sigma_pointsx2}, the columns of $\Lb$ are added and subtracted from the mean $\hat{\bxi}$ to form a set of $2n$ sigma points. 

 The sigma points \eqref{eq:Sigma_points} are then plugged into the nonlinear process model \eqref{eq:Nonlinear_Noise_f} ($\u_{k-1}$ is assumed known):
\begin{equation}\label{eq:pro_process}
 \hat{\bXi}^{(i)}_{k} = {\bf f} \big[{\bXi}^{(i)}_{k-1}, \u_{k-1}\big], \quad i = 0, 1, \ldots, 2n.
\end{equation}
The predicted mean $\hat{\bxi}_{k}$ and predicted covariance $\P_{k}$ are calculated as follows  \cite{Wan:ch7}:
\begin{align}\label{eq:predicted_state}
	\hat{\bxi}_{k } &= \sum_{i = 0}^{2n} w^{(i)}_m \hat{\bXi}^{(i)}_{k},\\ \label{eq:Pk}
	{\bf P}_{k }^{\bxi} &= \sum_{i = 0}^{2n} w^{(i)}_c\Big(\hat{\bXi}^{(i)}_{k} - \hat{\bxi}_{k}\Big)\Big(\hat{{\bXi}}^{(i)}_{k} - \hat{\bxi}_{k}\Big)^\T + \Q_{k-1},
\end{align}
where the weights $w^{(i)}_m$ and ${w}_c^{(i)}$ are given as follows:
\begin{subequations}\label{eq:Weights}
\begin{align}
{w}_m^{(0)} &= \frac{\lambda}{\lambda+n},\\
{w}_c^{(0)} &= \frac{\lambda}{\lambda+n} + 1 - a^2 + \epsilon,\\
{w}_m^{(i)} &\equiv {w}_c^{(i)} = \frac{1}{2(\lambda+n)}, \quad\quad i = 1, \ldots, 2n
\end{align}
\end{subequations}
and $\epsilon$ (for Gaussian distributions, $\epsilon =2$ is optimal) is a constant that incorporates prior information on the probability distribution of $\bxi$. 

{\it 2) Update at $k$ for $\bxi_{k}$}. Given the predicted mean $\hat{\bxi}_{k}$, additional sigma points can be obtained from the matrix square root of the noise covariance $\Q$ in the physical system \eqref{eq:Nonlinear_Noise_f} as follows:
\begin{subequations}\label{eq:Sigma_points1}
\begin{align}
\bXi^{(0)}_{k} &= \hat{\bxi}_{k} \in \Rn^n, \\
\bXi^{(i)}_{k} &= \hat{\bxi}_{k} + \delta \left[\sqrt{{\bf Q}_{k-1}} \right]_i, \quad i = 1, \ldots, n\\
\bXi^{(i+n)}_{k} &= \hat{\bxi}_{k} - \delta\left[\sqrt{{\bf Q}_{k-1}} \right]_i, \quad i = 1, \ldots, n
\end{align}
\end{subequations}
Alternatively, a new set of sigma points can be redrawn using the current (predicted) covariance, ${\bf P}_{k}^{\bxi}$ in \eqref{eq:Pk}: 
\begin{equation}\label{eq:Alt_SigmaPoints}
\bXi_{k } = 
\Big[
\hat{\bxi}_{k}^\T  \,\,\, \big(\hat{\bxi}_{k} +\delta \left[\sqrt{{\bf P}_{k}} \right]_i\big)^\T    \,\,\,  \big(\hat{\bxi}_{k} - \delta \left[\sqrt{{\bf P}_{k}} \right]_i\big)^\T 
\Big]^\T
\end{equation}
The various weights $w$ are also recalculated accordingly by setting $n \to 2n$.

\begin{algorithm}[t]
\caption{The UKF Algorithm.}\label{alg:cap}
\resizebox{\columnwidth}{!}{%
\begin{tabular}{|l|}\hline
{\bf Initial Conditions:} Mean and covariance of $\bxi_{k-1}$: $\hat{\bxi}_{k-1}$, ${\bf P}_{k-1}^{\bxi} = \I$\\
{\bf Output:} A posteriori state estimate: $\hat{\bxi}_{k | k}$ \\ \hline\hline
{\bf Prediction at $k-1$ for $\bxi_{k-1}$:}\\[3pt] \hline\hline
{\it 1.} Generate Sigma Points ${\bXi}^{(i)}_{k-1}$ according to \eqref{eq:Sigma_points} and corresponding\\ \quad weights $w^{(i)}_c$, $w^{(i)}_m$, $i = 0, 1, \ldots, 2n$, according to \eqref{eq:Weights}.\\
{\it 2.} Propagate the sigma points through the process model \eqref{eq:pro_process}:\\
$\begin{array}{ll}
&\quad \quad\hat{\bXi}^{(i)}_{k} = {\bf f} \big[{\bXi}^{(i)}_{k-1}, \u_{k-1}\big],\; i = 0, 1, \ldots, 2n.
\end{array}$
\\
{\it 3.} Compute the predicted mean $\hat{\bxi}_{k}$ and covariance $\P_{k}^{\bxi}$:\\
$\begin{array} {ll} 
	&\quad\quad\hat{\bxi}_{k} = \sum_{i = 0}^{2n} w^{(i)}_m \hat{\bXi}^{(i)}_{k},\\
	&\quad\quad{\bf P}_{k}^{\bxi} = \sum_{i = 0}^{2n} w^{(i)}_c\Big(\hat{\bXi}^{(i)}_{k} - \hat{\bxi}_{k}\Big)\Big(\hat{\bXi}^{(i)}_{k} - \hat{\bxi}_{k}\Big)^\T + \Q_{k-1}.
\end{array}$
\\ \hline\hline
{\bf Update at $k$ for $\bxi_{k}$:}\\ \hline\hline
{\it 1.}  Generate sigma Points ${\bXi}^{(i)}_{k}$ according to \eqref{eq:Sigma_points1} or \eqref{eq:Alt_SigmaPoints}, and\\ 
\quad weights $w^{(i)}_c$, $w^{(i)}_m$, $i = 0, 1, \ldots, 2n$ according to \eqref{eq:Weights}.\\
{\it 2.} Propagate sigma points through the measurements model \eqref{eq:Prop_Meas}:\\
$\begin{array}{llll}
&\quad\quad \hat{\bf Y}^{(i)}_{k} = {\bf g} \big[{\bXi}^{(i)}_{k | k-1}\big],\; i = 0, 1, \ldots, 2n
\end{array}$
\\
{\it 3.} Compute the predicted mean $\hat{\y}_{k}$ and covariance $\P_{k}^{\y}$ of $\y_k$:\\
$\begin{array} {ll} 
	&\quad \quad\hat{\y}_{k} = \sum_{i = 0}^{2n} w^{(i)}_m \hat{\bf Y}^{(i)}_{k},\\
	&\quad \quad{\bf P}_{k}^{\y} = \sum_{i = 0}^{2n} w^{(i)}_c\Big(\hat{\bf Y}^{(i)}_{k} - \hat{\y}_{k}\Big)\Big(\hat{\bf Y}^{(i)}_{k} - \hat{\y}_{k}\Big)^\T + \R_{k}.
\end{array}$\\
{\it 4.} Compute the {\it cross-covariance} between $\bxi_k$ and $\y_k$:\\
$\begin{array} {ll} 
	&\quad \quad{\bf P}_{k}^{\bxi\y} = \sum_{i = 0}^{2n} w^{(i)}_c\Big(\hat{\bXi}^{(i)}_{k} - \hat{\bxi}_{k}\Big)\Big(\hat{\bf Y}^{(i)}_{k} - \hat{\y}_{k}\Big)^\T.
\end{array}$\\
{\it 5.} Compute the {\it Kalman gain} using the a priori covariance:\\
$\begin{array} {llll} 
	&\quad\quad \K_k = \Pb_{k}^{\bxi\y}(\Pb_{k}^{\y})^{-1}.
\end{array}$\\
{\it 6.} Compute the a posteriori state estimate and covariance matrix:\\
$\begin{array} {llll} 
	&\quad\quad \hat{\bxi}_{k | k} = \hat{\bxi}_{k | k-1} + \K_k (\y_k - \hat{\y}_{k}),\\
	&\quad\quad \P_{k | k}^{\bxi} = \P_{k | k-1}^{\bxi} -\K_k (\Pb_{k}^{\bxi\y})^\T.
\end{array}$\\
\hline
\end{tabular}
}
\end{algorithm}

The new sigma points are then plugged into the nonlinear measurements model \eqref{eq:Nonlinear_Noise_g}, which results in the transformed sigma points of the output model:
\begin{equation}\label{eq:Prop_Meas}
 \hat{\bf Y}^{(i)}_k = {\bf g} \Big[{\bXi}^{(i)}_k\Big], \quad\quad i = 1, \ldots, 2n.
\end{equation}

Given the transformed sigma points through the state and measurement models, $\hat{\bXi}_k$ and $\hat{\bf Y}_k$, respectively, the predicted mean, $\hat{\y}_{k}$, predicted covariance of the measurements,  $\P_{k}^{\y}$, and the cross-covariance between $\bxi_k$ and $\y_k$ can be obtained as follows:
\begin{align}
\hat{\y}_{k} &= \sum_{i = 0}^{2n} w^{(i)}_m \hat{\bf Y}^{(i)}_{k},\\
{\bf P}_{k }^{\y} &= \sum_{i = 0}^{2n} w^{(i)}_c\Big(\hat{\bf Y}^{(i)}_{k} - \hat{\y}_{k}\Big)\Big(\hat{\bf Y}^{(i)}_{k} - \hat{\y}_{k}\Big)^\T + \R_{k},\\
{\bf P}_{k}^{\bxi\y} &= \sum_{i = 0}^{2n} w^{(i)}_c\Big(\hat{\bXi}^{(i)}_{k} - \hat{\bxi}_{k}\Big)\Big(\hat{\bf Y}^{(i)}_{k } - \hat{\y}_{k}\Big)^\T.
\end{align}
Finally, the filter gain $\K_k$, the a posteriori state (filtered) estimate and covariance matrix, conditional on the measurement $\y_k$ can be calculated:
\begin{align}
	\K_k &= \Pb_{k}^{\bxi\y}(\Pb_{k}^{\y})^{-1},\\
	\hat{\bxi}_{k | k} &= \hat{\bxi}_{k | k-1} + \K_k (\y_k - \hat{\y}_{k}),\\
	\P_{k | k}^{\bxi} &= \P_{k | k-1}^{\bxi} -\K_k (\Pb_{k}^{\bxi\y})^\T.
\end{align}
where both $\hat{\bxi}_{k | k-1}$ and $\P_{k | k-1}^{\bxi}$ obtained at $k -1$ from \eqref{eq:predicted_state} and \eqref{eq:Pk}, respectively.

\begin{figure}\centering
	\includegraphics[width=.95\columnwidth]{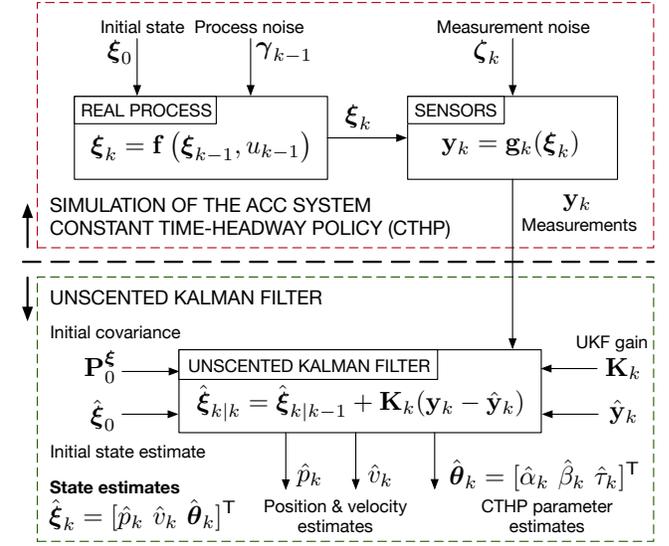}
\caption{CTHP parameter estimation based on unscented Kalman filter.}\label{fig:UKF_Scheme}
\end{figure}

Algorithm \ref{alg:cap} summarizes the main steps of the UKF, while Fig.\ \ref{fig:UKF_Scheme} illustrates the relationship between DUKF and ACC for CTHP parameter estimation.	

\section{Application and Results}\label{sec:results}

This section presents empirical data from a car-following experiment of an ACC engaged ego vehicle and a human-driven vehicle (leader). These data are then used to test the effectiveness of the proposed DUKF in identifying the CTHP parameters of a commercially implemented ACC system. The DUKF is also compared to batch and recursive least-squares optimization. Note that the proprietary control logic of the stock ACC controller and its true parameters are unknown.

\subsection{Empirical ACC Driving Data}

The empirical data of relative velocity and space headway were obtained from a real-life experiment conducted in the Autostrada A26 motorway in Italy (from Ispra to Casale Monferrato and vice versa) in 2020. The empirical data are freely available from the OpenACC repository\footnote{\url{http://data.europa.eu/89h/9702c950-c80f-4d2f-982f-44d06ea0009f}.} of the Joint Research Centre (JRC), European Commission \cite{MAKRIDIS2021103047}. The trial involved two vehicles in car-following formation in actual motorway driving conditions. The two vehicles  are a hydrogen fuel cell electric powered crossover SUV (Hyundai Nexo, 2019) and a mid-size diesel SUV (SsangYong Rexton, 2019).

During the experiment, the car-following order was the same, the driver of the lead vehicle  (SsangYong Rexton) was instructed to drive as they would normally in traffic, and the follower vehicle  (Hyundai Nexo) was driving at all times with the ACC engaged with minimum settings (i.e., the setting that allows the ACC ego vehicle to follow closest to the vehicle ahead). ACC disengagement happened only when the driver needed to manually brake, or the vehicle passed the minimum operating speed of the ACC. Also no overrides and no cut in behavior between the leader and the ACC follower occurred.

\begin{figure}\centering
	\includegraphics[width=\columnwidth]{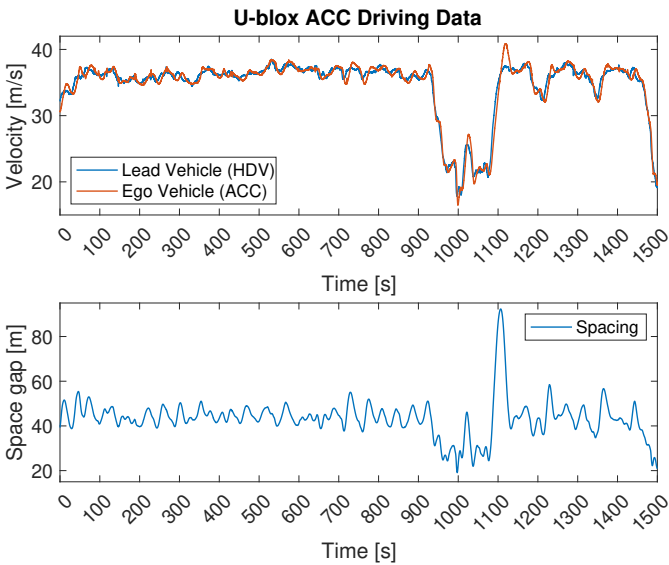}
\caption{Empirical data from a car-following experiment of an ACC engaged vehicle (Hyundai Nexo, 2019) and a human-driven vehicle (SsangYong Rexton, 2019).}\label{fig:data}
\end{figure}

To collect accurate data, both vehicles were equipped with U-blox M9 precision Global Navigation Satellite System (GNSS) units that track global location and velocity, and on-board diagnostics (OBD). Data were recorded at 10 Hz (0.1 s). Fig. \ref{fig:data} depicts the recorded observations of  relative velocity and space headway from the car-following system. The data contains both equilibrium and non-equilibrium driving conditions, which are used to identify the ACC parameters of the Hyundai Nexo using the proposed DUKF.

\subsection{DUKF Setup \& Parameter Estimation on a 2019 SUV}

To run the DUKF, the state model \eqref{eq:discrete_matrix} and measurement model \eqref{eq:meas} are corrupted by the process noise $\bm{\gamma}$ and measurement noise $\bm{\zeta}$, respectively, to agree with \eqref{eq:Nonlinear_Noise}.  Although the vector of parameters $\bm{\theta} = [ \alpha\,\, \beta\,\, \tau]^\T$ is assumed to be constant in \eqref{eq:discrete_matrix}, a small pseudo noise term is added, which may speed up the convergence of parameter estimates.    The DUKF ran with initial covariance matrix ${\bf P}_0 = \bf I$, and covariance matrices ${\bf Q} = {\rm diag}$(2.0e-05, 5.0e-06,   1.0e-06   1.0e-06  1.0e-06) and $\R={\rm diag}(0.8,0.2)$ for the physical system and measurement noise, respectively. The initial augmented state and its estimate are $\bm{\bxi}_0 = [40\; 30\; 0.1\; 0.2\; 1.2]^\T$  and $\hat{\bm{\bxi}}_0 = [ 35\;\; 25\;\;  0.08\;\; 0.12\;\; 1.5]^\T$, respectively. Note that the known set of design parameters, $\bm{\theta}_0 = [\alpha_0\,\, \beta_0\,\, \tau_0]^\T = [0.1\,\, 0.2\,\, 1.2]^\T$ and its estimate $\hat{\bm{\theta}}_0 = [\hat{\alpha}_0\,\, \hat{\beta}_0\,\, \hat{\tau}_0]^\T = [0.08\;\; 0.12\;\; 1.5]^\T$, corresponds to a string unstable system in terms of both the $\Ls_2$ and $\Ls_\infty$ norms, cf.\ with conditions \eqref{eq:L2c} and \eqref{eq:Linfc}, respectively. The unscented transform ran with $a = 1$, $b=3-n$, $\epsilon = 0$.

For the assessment of the ACC model parameter estimation, the {\it mean absolute error} (MAE)  (of the car-following system) in space-gap and velocity of the ACC engaged ego vehicle (Hyundai Nexo) is considered to assess the accuracy of the DUKF.  Since the actual parameters of the ACC engaged vehicle are unknown, the choice of the two empirically observed state variables (spacing and velocity) for assessing the DUKF is sound. The string stability of the calibrated CTHP of the stock ACC system under the parameters estimated by the DUKF is also calculated and reported.

Table \ref{tbl:DUKF_results} summarizes the obtained results for three different initial conditions of the filter. Each experiment runs 50 times and the average values are reported. In all cases, the reported ACC model parameters fit the data well (see the MAE values), albeit with some differences in the actual parameter values. The estimated time-headway of the ACC engaged ego vehicle (Hyundai Nexo) is in the bracket $\tau \in [1.1223, 1.1639]$ s, which is consistent with the  median value of the time headway obtained from the empirical data. The results are consistent for different initial conditions of the DUKF, provided that tracking of spacing and velocity profiles are accurate, i.e., their MAE is sufficient small and in scale to previous  studies considering commercial ACC systems \cite{MILANES2014,Gunter:2020,Gunter:2021,Wang:2021}.

\begin{table}\caption{Performance of the DUKF.}\label{tbl:DUKF_results}
\centering
\begin{tabular}{l|ccc}\hline\hline
Experiment & \#1 & \#2 & \#3 \\ \hline\hline
Estimated  & $\alpha = 0.1987$ & $\alpha = 0.1454$ & $\alpha = 0.2134$ \\
parameter & $\beta = 0.1294$ & $\beta = 0.1809$ & $\beta = 0.1849$\\
values & $\tau = 1.1639$ & $\tau = 1.1223$ & $\tau =  1.1305$\\ \hline
MAE space-gap (m) & 1.27e-01 & 1.36e-01 & 1.16e-01\\
MAE velocity (m/s) & 4.57e-02 & 4.77e-02 & 3.89e-02\\ \hline
$\Ls_2$ strict string stable & NO & NO & NO\\ 
$\Ls_\infty$ strict string stable  & NO & NO & NO\\ \hline
\end{tabular}
\end{table}

Given the estimated ACC model parameters by the DUKF in Table \ref{tbl:DUKF_results}, string stability is checked using \eqref{eq:L2c} and \eqref{eq:Linfc} conditions for the ACC engaged ego vehicle. As can be seen, the commercially implemented ACC system of the considered vehicle is neither $\Ls_2$ nor $\Ls_\infty$ strict string stable. This result is consistent to previous works  on the parameter identification of commercial ACC systems \cite{Knoop,Gunter:2021,Wang:2021}. 

\begin{figure}\centering
	\includegraphics[width=.98\columnwidth]{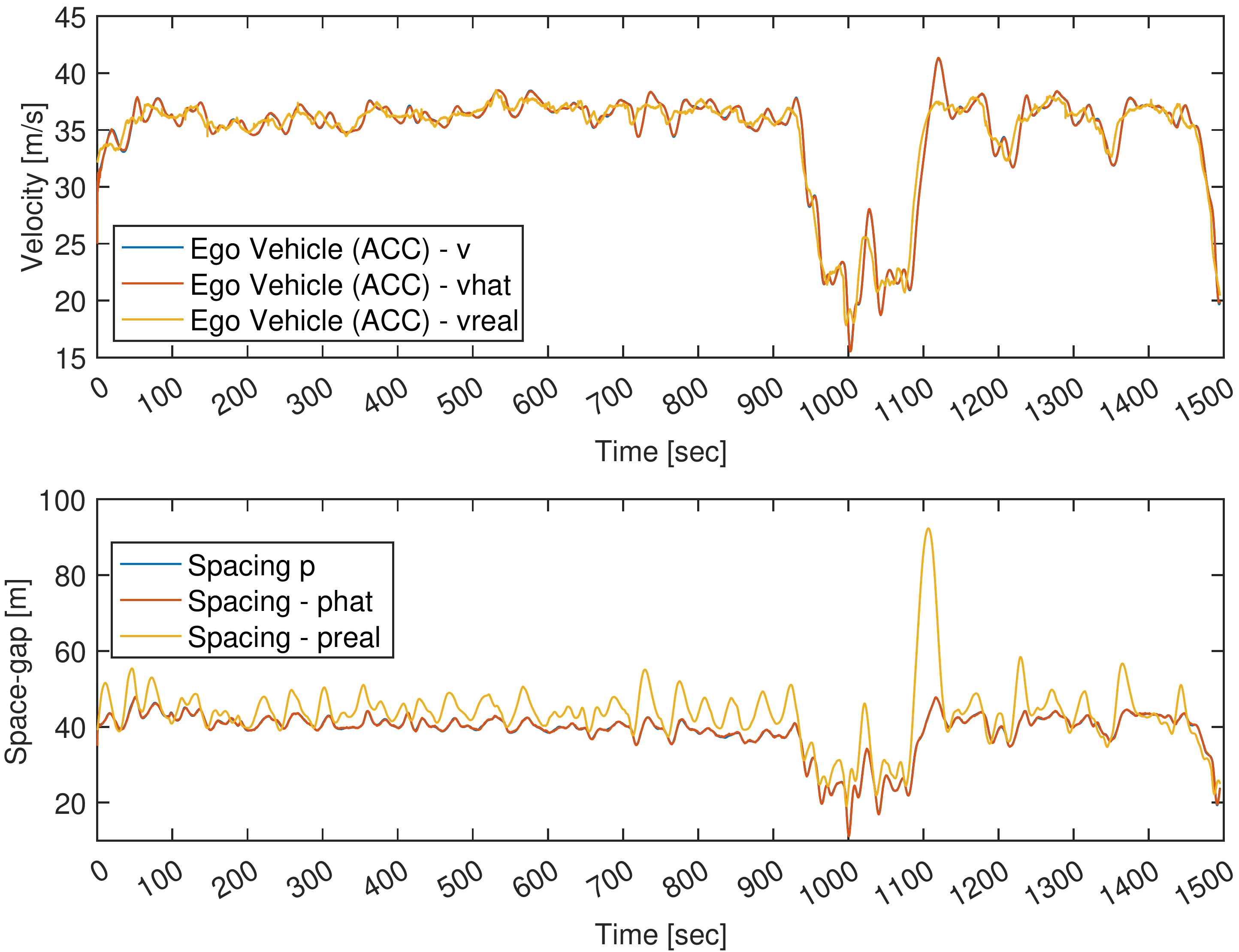}\\[5pt]
	\includegraphics[width=\columnwidth]{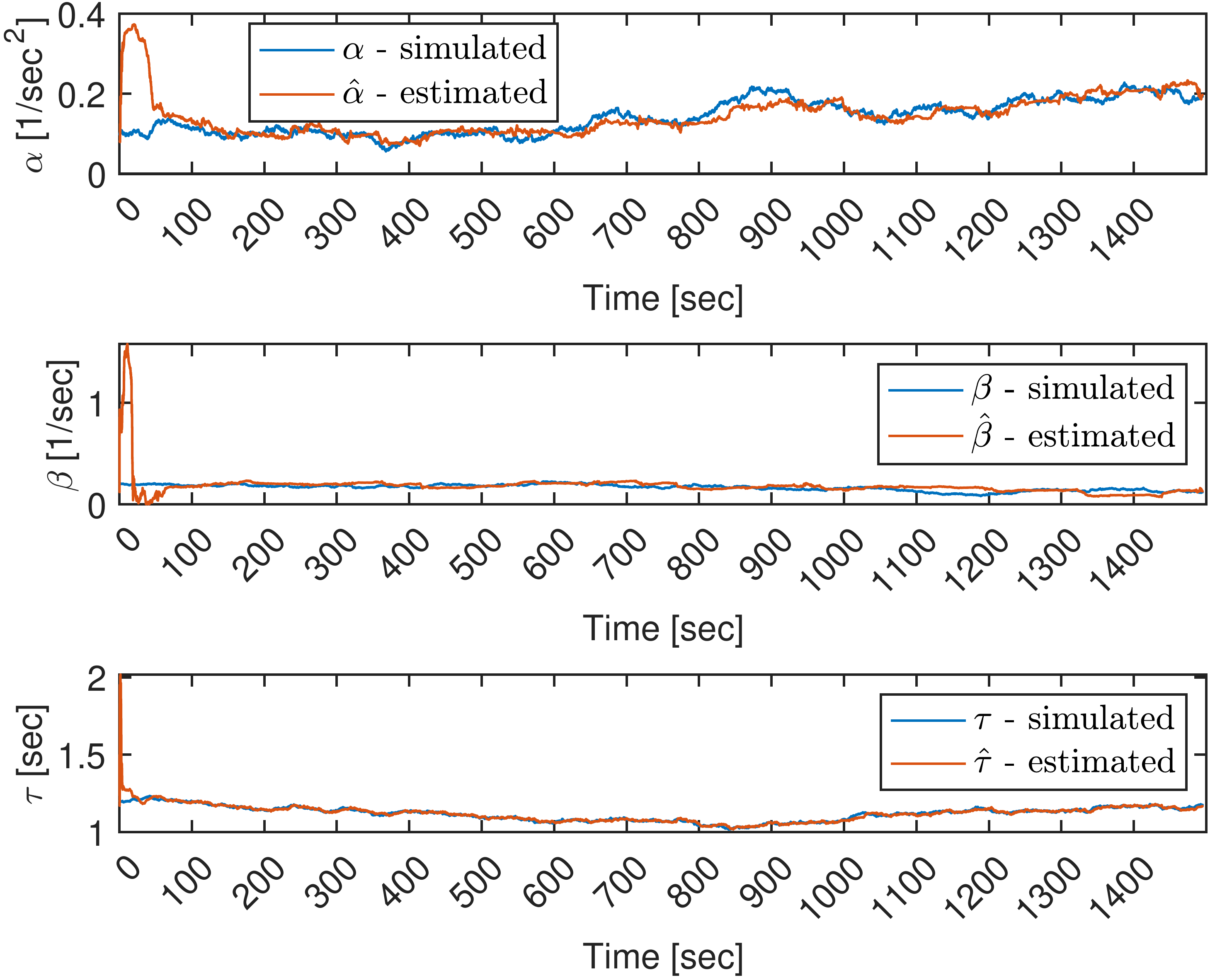}\\
\caption{Real-time estimation of the velocity of  the ACC ego vehicle, spacing, and ACC model parameters by the DUKF.}\label{fig:results}
\end{figure}

Fig.\ \ref{fig:results} depicts the obtained real-time estimates of the velocity of the ACC ego vehicle and the spacing between the two vehicles of the car-following system. These trajectories correspond to the ACC model parameters reported in Table \ref{tbl:DUKF_results} for $\alpha = 0.1987$ [1/s$^2$],  $\beta = 0.1294$ [1/s], and $\tau = 1.1639$ [s]. As can be seen, the DUKF delivers excellent tracking of the velocity profile of the ACC ego vehicle while spacing is slightly less tracked. This is in agreement with the MAE values reported in Table \ref{tbl:DUKF_results}. The largest error between the measured spacing and the CTHP estimated parameters arises between 900 s and 1100 s. In this time windows the real ACC vehicle in the trial engages in an acceleration that is not reproduced well by the DUKF. The last three subfigures of Fig.\ \ref{fig:results} depict the real-time estimates of the stock ACC model parameters. As can be seen,  the convergence of $\beta$ and $\tau$ is fast, while $\alpha$ is more agile and sensitive to non-equilibrium traffic conditions in which the real ACC vehicle in the trial engages in acceleration. This is attributed to the role of $\alpha$ in the constant time-headway policy \eqref{eq:CTHP}, which is to control the desired gap.  

\subsection{Comparison with Batch and Recursive Least-Squares}

This section compares the proposed DUKF with three other estimation approaches that are based on least-squares optimization; namely least-squares batch estimation  (LS-BE), least-squares recursive estimation  (LS-RE), and least-squares recursive estimation with exponential weighting (LS-REXP). For details of each approach see the Appendix \ref{ap:LS}.

Provided that real-time data of $v_k$, $u_k$, and $p_k$ are available, the discretized version of the CTHP \eqref{eq:CTHP} can be rewritten as: 
\begin{equation*}
	v_{k+1} = x_1 v_k + x_2 u_k  + x_3 p_k,
\end{equation*}
with $x_1 = 1 - (\alpha \tau + \beta)T$, $x_2 =  \beta T$, and $x_3 = \alpha T$; where now the vector of model parameters to be estimated is $\x = [ x_1\,\, x_2\,\, x_3]^\T \in \Rn^3$. Obviously, the CTHP model parameters, $\alpha$, $\beta$, and $\tau$, can be recovered once $\x$ is estimated via least-squares optimization according to the Appendix \ref{ap:LS}. 

The observation model \eqref{eq:obs} can be obtained using as an input a dataset of measurements ${\cal X}_{\kappa} = \{v_k, p_k, u_k\}$ for $k =0, 1,\ldots, \kappa-1 \le l$, with $l \ge 3$,
\begin{equation}\label{eq:LinearSystem}
\begin{bmatrix}
v_1\\ v_2\\ \vdots\\ v_{\kappa}
\end{bmatrix} = 
\begin{bmatrix}
v_0 & u_0 & p_0\\ 
v_1  & u_1 & p_1\\ 
\vdots & \vdots & \vdots\\ 
v_{\kappa-1}  & u_{\kappa-1} & p_{\kappa-1}
\end{bmatrix} 
\begin{bmatrix}
x_1\\ x_2\\  x_3
\end{bmatrix}, \text{ or } \, \z = \H\x,
\end{equation}
with the vector $\z \in \Rn^\kappa$ comprising the values of $v_k$ for $k=1, 2, \ldots, \kappa$, and the matrix $\H \in \Rn^{\kappa \times 3}$ comprising the dataset ${\cal X}_{\kappa}$, for $k=0, 1, \ldots, \kappa-1$. The linear system \eqref{eq:LinearSystem} obeys a unique solution if and only if ${\rm rank}(\H) = 3$.

The least-squares batch estimator  (LS-BE)  \eqref{eq:LS} is setup with $\R = \I$ and $\sigma =0.001$. The least-squares recursive estimator  (LS-RE) and recursive estimator with exponential weighting (LS-REXP) run with initial conditions $\x_0 = \big[0.98\,\, 0.01\,\, 0.01\big]^\T$ (corresponding to $\alpha_0 = 0.1$, $\beta_0 = 0.1$, $\tau_0 = 1$),  $\P_0 = 0.001\times \I_3$, and exponential weighting factor $\mu = 1.01$ (i.e., future values of the measurements are slightly more important than past values to improve learning). All three least-squares estimators fed with the recorded observations of  relative velocity and space headway depicted in Fig. \ref{fig:data}. These data contain both equilibrium and non-equilibrium driving conditions.

Table \ref{tbl:LS_results} summarizes the obtained results for the three least-squares estimators. As can be seen, for the one-shot LS-BE the parameter $\beta$ takes a negative value. This is possible since the least-squares estimator is unconstrained. The LS-REXP (recursive with exponential weighting) achieves both the lowest MAE velocity and space gap errors at $0.34$ m/s and $2.23$ m. The LS-RE (recursive) method has a comparable performance, with MAE values of $0.42$ m/s and $3.21$ m. Overall, the MAEs are comparable to those found in other works employing least-squares estimation, see e.g.\ \cite{Gunter:2021,Wang:2021}. 

Comparing with Table  \ref{tbl:DUKF_results}, the proposed DUKF is seen to be always better  (see the MAE values) than all three versions of the least-squares optimization. This is attributed to the fact that the proposed DUKF delivers parameter estimates for the nonlinear dynamics of the CTHP in the presence of measurement noise, while least-squares optimization is data-driven without full knowledge of the state-space model. Finally in Table \ref{tbl:LS_results}, all models estimated via least-squares optimization under the estimated parameters are seen to be $\Ls_2$ and $\Ls_\infty$ strict string unstable.

\begin{table}\caption{Performance of Least-Squares Estimation (Appendix \ref{ap:LS}).}\label{tbl:LS_results}
\centering
\begin{tabular}{l|ccc}\hline\hline
Experiment & LS-BE & LS-RE & LS-REXP \\ \hline\hline
Estimated  & $\alpha = 0.0062$ & $\alpha = 0.0042$ & $\alpha = 0.0125$ \\
parameter & $\beta = -0.1143$ & $\beta = 0.0969$ & $\beta = 0.0819$\\
values & $\tau = 1.2801$ & $\tau = 1.2750$ & $\tau =  1.2946$\\ \hline
MAE space-gap (m) & 4.59 & 3.21 & 2.23\\
MAE velocity (m/s) & 0.84 &  0.42 & 0.34\\ \hline
$\Ls_2$ strict string stable & NO & NO & NO\\ 
$\Ls_\infty$ strict string stable  & NO & NO & NO\\ \hline
\end{tabular}
\end{table}

\section{Conclusions and Outlook}\label{sec:conclusions}

This paper developed and investigated a dual unscented Kalman filter for the joint state and parameter identification of commercially implemented ACC systems using empirical data from a real-life car-following experiment. For the ACC system, a constant time-headway policy was considered, and its parameters were considered to be unknown. The set of ACC model parameters obtained from the proposed estimation scheme for the particular CTHP revealed that the commercially implemented ACC system of a Hyundai Nexo SUV (2019) is neither $\Ls_2$ nor $\Ls_\infty$ string stable. However, the controller type of the particular vehicle and its parameters are not publicly available, so any conclusions must be drawn with caution.

The nonlinear ORC presented in Section \ref{sec:NonLinORC} is amenable to the type of the adopted discretization scheme and the reference input signal (leader's velocity), so it may be fulfilled under certain conditions, e.g., if the reference input signal satisfies a persistent excitation condition \cite{BOYD1986_PE} or a better discretization scheme is employed to approximate the continuous-time dynamics in \eqref{eq:p}--\eqref{eq:CTHP}. This would be an avenue for future research.

Despite criticisms, commercially implemented ACC systems are likely to improve in the near future using enhanced connectivity and cooperation via V2X (Vehicle-to-Vehicle and Vehicle-to-Infrastructure) communication to ensure safe and quick response to perturbation events much further downstream in a promptly manner, and thus, generating smoother responses. In conclusion, with enhanced connectivity, string stability might be achievable because V2V communications permit tighter vehicle spacing control, so that inter-vehicle time-gap settings are significantly shorter than the stock ACC time-gap settings.

Future work will address the parameter identification of cooperative ACC (CACC) and ADAS, including personalized driving, spacing policies  in the presence of parasitic actuator lags and connectivity delays, as well as variable time headway policies, using higher-order vehicle dynamics while  considering comfort and eco-driving instructions \cite{Apostolakis:BC_Energy}. The definition of \emph{input-to-output string stability} in Section \ref{sec:stability}  concerns systems as mappings between inputs and outputs, but it ignores  internal and external system disturbances (only the platoon leader is subject to external disturbances).  \emph{Input-to-state string stability} will be explored to explicitly consider the effects of initial perturbations and external disturbances on each vehicle on a platoon \cite{Besselink}.

\appendices

\section{Batch and Recursive Least-squares Estimation}\label{ap:LS}

Consider the problem of offline single-stage (batch) estimation (LS-BE) where measurements of a constant vector are being corrupted by noise:
\begin{equation}\label{eq:obs}
	\z = \H\x + \bm{\zeta},
\end{equation}
where  $\z \in \Rn^l$ is a vector of observations, $\x \in \Rn^n$ is a vector of parameters to be estimated, $\H \in\Rn^{l\times n}$ is a matrix with linearly independent columns (which implies $l \ge n$), and $\bm{\zeta} \in \Rn^l$ is some observation noise that is unknown, but presumed to be sufficient small.  In this setting there is no information on the probability distribution of $\x$ and $\bm{\zeta}$, and thus statistically-based estimators cannot be developed.

The best deterministic estimate $\hat{\x}$ such that the {\it Tikhonov's regularized} least-squares criterion ({\it ridge regression}),
\begin{equation}\label{eq:cost}
	J (\x)= \frac{1}{2}\| 	\z - \H\x \|^2_{\bf R^{-1}} + \frac{1}{2}\sigma\|\x\|^2
\end{equation}
is minimum, where $\bf R$ is a symmetric positive definite matrix and $\sigma > 0$ is a regularization   parameter, can be found by ordinary calculus as \cite{Golub},
\begin{equation}\label{eq:LS}
	\hat{\x} = \big[\H^\T\bR^{-1}\H + \sigma\I \big]^{-1}\H^\T\bR^{-1}\z.
\end{equation}
The regularization (or penalty) parameter  $\sigma$ gives a compromise between making $\|\z - \H\x \|$ zero and keeping $\x$ of reasonable size. Moreover, since $\H^\T\bR^{-1}\H + \sigma\I \succ \bf 0$ for any $\sigma > 0$, the Tikhonov regularized least-squares solution demands no rank assumptions on $\H$. This is particularly useful in cases where $\H$ is ill-conditioned, or even singular.  To obtain \eqref{eq:LS}, reliable and efficient algorithms such as the Schwarz-Rutishauser algorithm with computational complexity $O(n^2 l)$ for the QR factorization can be used.

It should be highlighted that the minimization of \eqref{eq:cost} is equivalent to maximizing the conditional probability $\mathcal{P}(\z | \x)$ subject to \eqref{eq:obs} where $\bm{\zeta}$ is white Gaussian with zero mean and covariance $\bR$. Maximization of $\mathcal{P}(\z | \x)$ results in the maximum likelihood estimator.

Consider now the case where $\hat{\x}_{k-1}$ is computed for $k\le l$ measurements via \eqref{eq:LS} and an additional measurement $\z_{k}$ is available. The correction $\Delta\hat{\x}_{k} := \hat{\x}_{k} - \hat{\x}_{k-1}$ is given by:
\begin{equation}\label{eq:correction}
\begin{aligned}
	\Delta\hat{\x}_{k} &= [\H^\T_{k}\bR_{k}^{-1}\H_{k} + \H^\T_{k-1}\bR^{-1}_{k-1}\H_{k-1}]^{-1}\\
					&\quad\,\,\, \H^\T_{k}\bR_{k}^{-1}(\z_{k} - \H_{k}	\hat{\x}_{k-1}).
\end{aligned}
\end{equation}
Here a recursive  estimation scheme allows for the determination of $\hat{\x}_{k}$ without the inversion of a possibly high-dimensional matrix in  \eqref{eq:correction}. This may be achieved by the use of the 
{\it Sherman-Woodbury-Morrison formula} (or matrix inversion lemma) in \eqref{eq:correction} \cite{Golub}.
\begin{lem}[Sherman-Woodbury-Morrison Lemma]\label{lem:MIL}
  Let $\A$ and $\B$ be square invertible matrices, and let $\C$ be a
  matrix of appropriate dimension. Then, if all the following
  inverses exist, it holds:
  \begin{equation*}
  [\A + \C \B \C^\T]^{-1} = \A^{-1} - \A^{-1} \C [\B^{-1} + \C^\T \A^{-1} \C]^{-1} \C^\T \A^{-1}.
  \end{equation*}
\end{lem}

Applying Lemma \ref{lem:MIL} to \eqref{eq:correction} yields the recursive estimation (LS-RE)  scheme,
\begin{equation}\label{eq:recursive}
	\hat{\x}_{k} = \hat{\x}_{k-1} + \P_{k} \H^\T_{k}\bR^{-1}_{k}(\z_{k} - \H_{k}\hat{\x}_{k-1})
\end{equation}	
\begin{equation}
	\label{eq:recursiveP1}
		\P_{k} = \P_{k-1}-\P_{k-1}\H_{k}^\T[\H_{k}\P_{k-1}\H_{k}^\T + \bR_{k} ]^{-1}\H_{k}\P_{k-1}
\end{equation}
with initial pseudo-inverse of the input data up to $k-1$,
\begin{equation}\label{eq:recursiveP2}
	\P_{k-1} = \big[\H^\T_{k-1}\bR^{-1}_{k-1}\H_{k-1} + \sigma\I \big]^{-1}.
\end{equation}
Therefore the new estimate in \eqref{eq:recursive} is equal to the old one plus a linear correction term based on the new observations and $\P_{k-1}$ only, see the recursive equation \eqref{eq:recursiveP1} with the initial condition \eqref{eq:recursiveP2}. Importantly (thanks to Lemma \ref{lem:MIL}), the quantity $[\H_{k}\P_{k-1}\H_{k}^\T + \bR_{k} ]$ is a scalar, and no matrix inversion is required in \eqref{eq:recursiveP1}. Since only one new measurement is available at each step, note that $\H_{k} \in\Rn^{1\times n}$ and $\bR_{k} \in\Rn$.

A final version of the  recursive least-squares estimation scheme can be obtained using the so-called {\it exponential weighting}. To this end, an exponential memory term $\mu^{k}$, where $k$ is the current step and $\mu$ is a positive parameter, is used in the least-squares cost criterion \eqref{eq:cost} to weight more or less future measurements. If $\mu > 1$ later values of the measurements are more important than earlier values; the opposite is true for $ \mu < 1$, in which case $\mu$ is called the discount or forgetting factor. The parameter update equation for recursive least-squares with exponential weighting is the same as in \eqref{eq:recursive} while the right-hand side of \eqref{eq:recursiveP1} must be multiplied by $1/\mu$.

\bibliographystyle{IEEEtran}
\bibliography{IEEEabrv,UKF}

\begin{IEEEbiography}[{\includegraphics[width=1in,height=1.25in,clip,keepaspectratio]{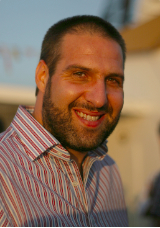}}]{Konstantinos Ampountolas}
(Member, IEEE) received the Dipl.Ing. degree in production engineering and management, the M.Sc. degree in operations research, and the Ph.D. degree in engineering from the Technical University of Crete, Greece, in 1999, 2002, and 2009, respectively.

He was a Senior Lecturer with the James Watt School of Engineering, University of Glasgow, U.K., from 2013 to 2019, a Research Fellow with the {\' E}cole Polytechnique F{\' e}d{\' e}rale de Lausanne, Switzerland, from 2012 to 2013, a Visiting Researcher Scholar with the University of California at Berkeley, Berkeley, CA, USA, in 2011, and a Post-Doctoral Researcher with the Centre for Research \& Technology Hellas, Greece, in 2010. He was also a short-term Visiting Professor with the Technion–Israel Institute of Technology, Israel, in 2014, and the Federal University of Santa Catarina, Florianópolis, Brazil, in 2016 and 2019. Since 2019, he has been an Associate Professor with the Department of Mechanical Engineering, University of Thessaly, Greece. His research interests include control and optimization with applications to transport networks and systems.

He has served as the Editor for \emph{Transportation} of \emph{Data in Brief}, from 2018 to 2019, as an Associate Editor for the \emph{Journal of Big Data Analytics in Transportation}, from 2018 to 2020, and on the editorial advisory boards of \emph{Transportation Research Part C} (from 2014 to 2021) and \emph{Transportation Research Procedia} (since 2014).
\end{IEEEbiography}

\vfill

\end{document}